\documentclass{aa}  
\usepackage[]{graphicx}
\usepackage{hyperref}
\usepackage[dvipsnames]{xcolor}
\usepackage{changes}
\usepackage{orcidlink}
\usepackage{comment}

\usepackage{txfonts}
\usepackage{amsmath}
\bibpunct{(}{)}{;}{a}{}{,}
\usepackage{linenoaa}
\usepackage{newtxtt}

\linenumbers

\newcommand{\sna}{SN~Ia}

\newcommand{\msun}{\mbox{$\mathrm{M_{\odot}}$}}

\newcommand{\lsun}{\mbox{$\mathrm{L_{\odot}}$}}

\newcommand{\mdot}{\mbox{$\mathrm{\dot{M}}$~}}
\newcommand{\mdote}{\mbox{$\mathrm{\dot{M}}$}}
\newcommand{\msyrm}{\mbox{$\mathrm{M_{\odot} yr^{-1}}$}}
\newcommand{\isotope}[2]{${}^{#1}$#2}

\newcommand{\mwd}{\mbox {{\rm M$_{\rm WD}$}}}

\newcommand{\rwd}{\mbox {{\rm R$_{\rm WD}$}}}
\newcommand{\gcc}{\mbox {{\rm g$\cdot$cm$^{-3}$}}}

\newcommand{\ptf}{PTF~J2238}
\newcommand{\cdth}{CD-30$^\circ$11223}
\newcommand{\myr}{\mbox {~${\rm M_{\odot}~yr^{-1}}$}}

\begin{document}

   \title{Will a Supernova explode in \cdth?\\ {\it In memory of Amedeo Tornamb\'e}}

   \author{L. Piersanti\orcidlink{0000-0002-8758-244X}\inst{1,2}
        \and
          L.R. Yungelson \orcidlink{0000-0003-2252-430X} \inst{3}
        \and
          E. Bravo \orcidlink{0000-0003-0894-6450}\inst{4}
        \and
          I. Dom{\'\i}nguez \orcidlink{0000-0002-3827-4731}\inst{4}\thanks{I. Domingues deceased during the writing of the present manuscript.}
          }

   \institute{INAF-Osservatorio Astronomico d'Abruzzo, via Mentore Maggini, snc, I-64100, Teramo, Italy\\
              \email{luciano.piersanti@inaf.it}
         \and
             INFN - Sezione di Perugia, via Pascoli, 123, I-06123, Perugia, Italy
         \and
             Institute of Astronomy of the Russian Academy of Sciences, 48 Pyatnitskaya Str, 119017 Moscow,
             Russia\\
             \email{lev.yungelson@gmail.com}
         \and
             Departamento de F\'{i}sica Te\'{o}rica y del Cosmos, Universidad de Granada, E-18071 Granada, Spain \\
             \email{eduardo.bravo@ugr.es}
             }

   \date{Received ; accepted }

  \abstract
   {Accretion of He-rich matter onto a low-mass carbon-oxygen white dwarf in a binary system with a He-rich donor may lead to an explosive event of Supernova Ia proportion, though with low peak luminosity and peculiar nucleosynthesis. Recently such a possibility for some binaries with He-accreting white dwarfs has been questioned, suggesting that, if the effects of rotation are accounted for in the evolution of the accretor, the latter does not explode, but it becomes a white dwarf with a massive He-buffer.} 
   {We investigate the expected evolution of the currently detached binary \cdth\ harboring a 0.74\,\msun\ carbon-oxygen white dwarf and a 0.47\,\msun\ donor with a He-burning core and a very thin H-envelope ($\Delta M_H=6\times 10^{-4}$\,\msun).}
   {We use the stellar evolution code FuNS to compute the evolution of \cdth\ up to the epoch when the two components come into contact and through the RLOF phase of the donor. With respect to our earlier study of the system PTF J2238+743015.1, we also include the transport of angular momentum due to magnetic instabilities in the accretor interiors (``magnetic model'').}
   {During the H-accretion phase, the effects of rotation in the accretor are negligible. In the ``magnetic model'', the angular momentum deposited by the accreted matter is very efficiently redistributed along the whole WD due to magnetic instabilities, so that the angular velocity of the accreted layers remains very low. During the He-accretion phase, the accretor experiences two very strong He-flashes, which result in the ejection from the binary system of a large part of the matter previously accreted. The system ends its life as a carbon-oxygen core capped by a massive He/C/O-envelope ($\Delta M_{\rm env}\simeq 0.194$\,\msun) and an extremely low-mass companion, remnant of the donor, like in the ``rotating model'' not including the effects of magnetic field.}
   {The system \cdth\ cannot be regarded as the potential progenitor of Supernova Ia. Such a conclusion applies also to all detached binary systems having similar masses of components and orbital periods.} 

\authorrunning{L. Piersanti et. al.}

\titlerunning{Will a Supernova explode in \cdth?}
   \maketitle
   \nolinenumbers

\section{Introduction}\label{s:intro}

The close binary system \cdth\ (GALEX J1411-3053), henceforth CD-30, hosting a carbon-oxygen (CO) white dwarf (WD) and a hot type B subdwarf (sdB) companion, was discovered by \citet{2012ApJ...759L..25V}.
\citet{2024MNRAS.527.2072D} derived the ``preferable'' set of the system orbital parameters $P_{orb}=70.53$\,min., $M_{\rm sdB}=0.47^{+0.07}_{-0.06}$\,\msun, $M_{\rm WD}=0.74^{+0.02}_{-0.02}$\,\msun. 
CD-30 belongs to a small sample of sdB+WD binaries \citep[see Tab. 2 in ][]{2025A&A...704A..82R} with known orbital periods and masses of components.
These systems are important, first of all, because 
it is deemed that, after Roche lobe overflow (RLOF) by the subdwarf and accretion of a certain amount of helium onto the WD, detonation of He may happen in the accreted layers, at least in some of these systems \cite{1987ApJ...317..717I}. 
Modern scenarios envision that the propagation of converging shock waves to the WD center may result in detonation of carbon, which produces a \sna\ scale event  \citep{Livne1990,lg91}.
On the other hand, depending on the combination of masses of components in a specific binary, a merging double-degenerate system may form. 
Recent studies of the outcome of He-detonation \citep{2021ApJ...919..126B,Shen_2021} demonstrated that a He-layer less massive than $\sim 0.03$\,\msun~ accreted onto a CO WD more massive than 0.9\,\msun~ is large enough to produce an explosion closely resembling a typical \sna~ event \citep[see also ][for a review]{2023ApJ...951...28W}. Conversely, in a typical low-mass CO WD \mwd$\le$ 0.8\,\msun\ the He-buffer should be very large ($\Delta M_{He}\ge$ 0.1\,\msun) to trigger a successful detonation \citep{lt91,ww94,hillman2026}. In the latter case, in the original He-buffer isotopes like \isotope{44}{Ti}, \isotope{48}{Ti}, and \isotope{51}{V} are largely overproduced as compared to a normal \sna~, the outer layers of the ejecta are dominated by high velocity Ni and unburnt He, and the associated light curve results dimmer \citep[see for instance][]{ww94,la95}.

\citet{Kupfer2022} and \citet{2024MNRAS.527.2072D} modelled the evolution of two detached sdB+WD systems from the above-mentioned sample --  PTFJ2238+743015.1 (hereinafter \ptf) and CD-30, respectively -- assuming that the WD components in them do not rotate.
They found that, in both cases, upon RLOF by the subdwarf 
and after consumption of its H envelope, accretion of He onto the WD results in a double-detonation and a weak \sna\ with peculiar nucleosynthesis.
On the contrary, \citet[][hereinafter Paper I]{piersanti2024}, who modelled \ptf\ assuming that the WD is rotating and accounting for shear heating, have shown that the WD experiences a series of very strong non-dynamical He-flashes. After that, due to the continuous decrease of the mass transfer rate, the reduced He-abundance in the accreted matter from the donor and the reduced compressibility of the accreted layers determined by the rotation-driven decrease of the local gravity, physical conditions suitable for He-burning in the He-mantle over the CO core cannot be realized. As a result, they conclude that in PTF-like systems He-detonation can not occur at all if the effects of rotation are accounted for in the evolution of the accreting WD.
Basing on a grid of models of sdB+non-rotating WD systems evolved up to He-ignition, \citet{2025A&A...704A..82R} favor a strong first He-flash in this system but do not exclude the possibility of a double detonation.  

In the present work, we extend the study of sdB+WD systems to CD-30, assuming that the WD is rotating and that angular momentum is redistributed inside the accreting WD due to the Tayler-Spruit instability.

The paper is structured as follows. 
In Section \ref{s:numerics}, we describe our physical assumptions and numerical method.
In Section \ref{s:initial}, the accepted parameters of the system are presented.
Section \ref{s:no-rot} provides results of computations made under the assumption that the accreting WD does not rotate, 
while in Section \ref{s:rot} we display the results for the rotating model without the inclusion of magnetic effects.
In Section \ref{s:ts}, the model accounting for the Tayler-Spruit instability is considered.
Finally, in Section \ref{s:concl}, the results of the study are discussed and summarized.

\section{Input Physics and Numerics}\label{s:numerics}
All models have been computed with the FuNS code \citep{straniero2006,cristallo2009,piersanti2013}. 
We recall that this is a 1D implicit Lagrangian hydrostatic code and, hence, the effects of rotation on the physical properties of a star are accounted for according to the method described in \cite{endal78}\citep[see the discussion in Sect. 2 of ][]{piersanti2013}. Angular momentum transport is treated as a pure diffusive process by assuming that rigid rotation is enforced in convectively unstable zones (see the discussion in Sect. 2 of \cite{piersanti2013} and in Sect. 2 of Paper I). The mixing of matter inside a star as determined by convection and rotation-induced instabilities is modeled as a diffusive process as detailed in \citet[][- see also Paper I]{piersanti2013}.
With respect to our previous study, we consider also the rotation-induced magnetic instability. To this aim we adopt for the Tayler-Spruit mechanism the formalism introduced by \citet{spruit2002}, as summarized in Sect. 2.1.1 of \citet[][Equations (4) - (12)]{neunteufel2017}.
For consistency with the adopted physical inputs in our code, we compute the magnetic diffusivity, necessary to implement the angular momentum transport and related mixing efficiencies, according to the prescription in \citet[][ and references therein]{itoh1987}.
All other physical assumptions are the same as in Paper I. In particular in our models we do not account for gravitational settling, diffusion, and levitation. As in Paper I \citep[see also][]{pier2014,piersanti2015,piersanti2019} we assume that the matter lost from the WD via RLOF during flash episodes is ejected from the binary system with the specific angular momentum of the accretor.

Finally, if during the accretion process the accretor attains the critical rotation rate on the surface, we assume that matter continues being accreted while the deposition of angular momentum is allowed at the level supporting critical rotation only \citep{pacz1991,popham1991,1993A&A...274..796B}.
\begin{figure}[t]
	\centering
	\includegraphics[width=0.8\columnwidth]{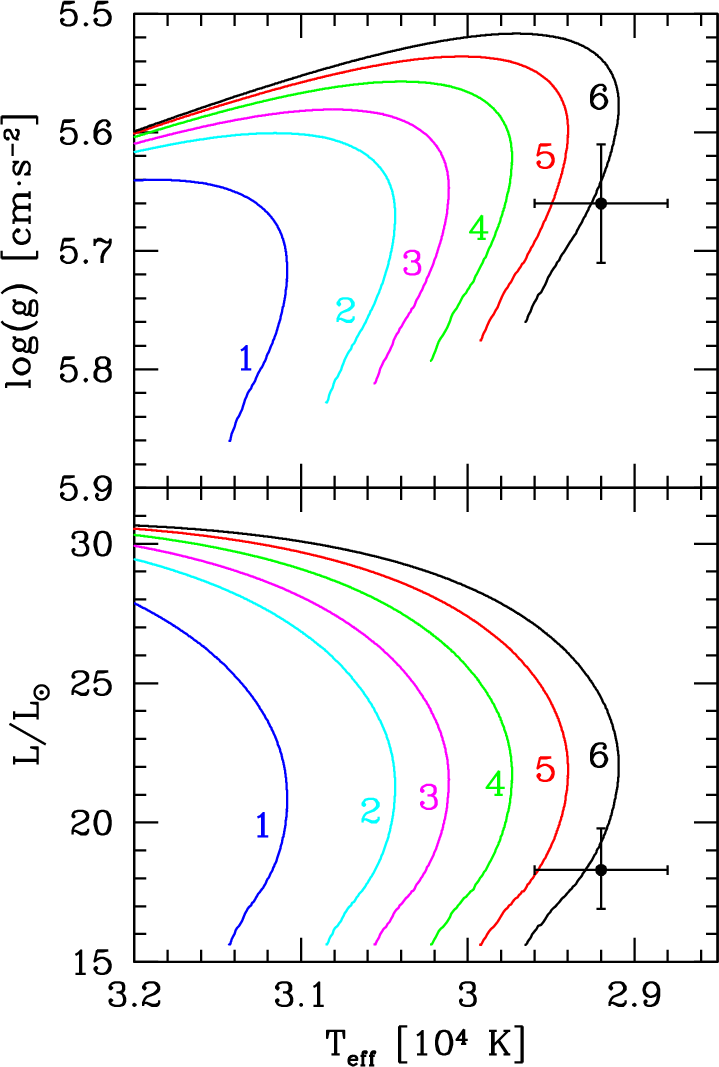}
	\caption{Surface gravity (upper panel) and luminosity (lower panel) as a function of the effective temperature of sdB models computed with fixed total mass (0.47\,\msun) and different masses of the H-rich skin, as reported in the plot (in $10^{-4}$\,\msun\, units). Points with error bars have been derived from \citet{2024MNRAS.527.2072D}, see their Tab. 1.}
	\label{fig1}
\end{figure}

\section{The initial system}\label{s:initial}
Based on the analysis by \citet{2024MNRAS.527.2072D}, we model the current CD-30 system as a binary harboring a CO WD of 0.74\,\msun\ and a 0.47\,\msun\ sdB star burning He in the core, with an orbital period $P_{orb}=70.53$\,min. 
The CO WD initial model has been constructed, as in Paper I, by accreting He-rich matter onto the ``heated'' model M070 from \citet{pier2014} at $\mdot = 3 \times 10^{-7}$\,\myr\ in order to simulate the evolution during the AGB phase.
The initial sdB model has been obtained by computing a series of stellar structures burning He in the core with fixed total mass and varying the mass of the H-rich skin (see Fig.~\ref{fig1}), in order to reproduce the values of surface gravity, effective temperature, and luminosity currently observed for the sdB component. 
The best fit was obtained by adopting $6\times 10^{-4}$\,\msun\ for the mass of the H-rich skin. 
According to our analysis, the compact binary system formed 38.03 Myr ago with an orbital period of $P_0=88.17$\,min.

We assume that, soon after formation of the compact configuration, the rotation of both components becomes synchronized with the orbital motion.
As in Paper I, we assume that the evolution of the period and, hence, of the orbital separation is completely driven by gravitational wave radiation.
By computing the subsequent evolution of this system, we found that the sdB component fills its Roche lobe in 23.82 Myr. Afterwards, the system remains semidetached during the entire traced evolution \citep[see also][]{teckenburg2025}.
\begin{table}
	\caption{Selected properties of the non-rotating model during the H-accretion phase.} 
	\label{hflnor}      
	\centering                   
 	 \begin{tabular}{l c c c c }
	  \hline\hline                 
	   \# Fl.            &   1  &   2  &   3  &   4   \\ 
	  \hline
       $\Delta T$        & 3.88 & 4.76 & 7.30 & 2.73  \\
       $\rho_{ig}$       & 2.82 & 2.54 & 2.56 & 3.86  \\
       $T_H^{max}$       & 1.43 & 1.43 & 1.49 & 1.76  \\
       $L_H^{max}$       & 5.12 & 5.04 & 6.39 & 8.36  \\
       $\Delta M_{acc}$  & 1.72 & 1.69 & 1.75 & 2.25  \\
       $\Delta M_{lost}$ & 1.70 & 1.69 & 1.76 & 2.03  \\
       <\mdote>          & 4.44 & 3.55 & 2.40 & 8.23  \\
       $\eta$            & 0.01 & 0.00 & 0.0  & 0.10 \\
      \hline                 
	 \end{tabular}
	 \tablefoot{For each flash episode we report the duration of the accretion phase ($\Delta T$ in Myr), the ignition density ($\rho_{ig}$ in $10^3$\gcc), the maximum temperature attained in the H-burning shell ($T_H^{max}$ in $10^8$ K) and the maximum energy per unit time delivered via H-burning ($L_H^{max}$ in $10^6$\lsun) during the flash, the mass accreted between two successive flashes ($\Delta M_{acc}$ in $10^{-4}$\,\msun), the mass ejected during the RLOF episode ($\Delta M_{lost}$ in $10^{-4}$\,\msun), the average accretion rate before the onset of RLOF (<\mdote> in $10^{-11}$\msyrm), and the retention efficiency defined as $\eta=1-\Delta M_{lost}/\Delta M_{acc}$.}
\end{table}

\section{Non-rotating model}\label{s:no-rot} 

As in Paper I for \ptf, we first computed the evolution of the CD-30 system ignoring the effects of rotation.
We find that the accretor experiences four H-flashes followed by RLOF by its expanding envelope, during which the entire mass accreted before the flash is completely lost from the system.
After that, the WD experiences an additional H-flash, the last one, but due to the low H-abundance in the accreted matter, the energy injected via nuclear burning is not sufficient to trigger a new mass loss episode (see discussion in Paper I, Sect. 4).
In Table~\ref{hflnor}, we report selected properties of the system during the four H-flashes followed by RLOF episodes.
The number of flashes may be compared to the twelve flashes in the non-rotating model for \ptf.
Such a difference arises mainly because, in \ptf, the mass of the H-rich layer on the sdB component is $2.1\times 10^{-3}$\,\msun\ (see Sect. 3 in Paper I), a factor 3 larger than the one in CD-30. The slight differences in the orbital parameters (masses of the WD and sdB and period at the onset of the very first mass transfer episode) have a second-order effect in determining the number of H-flashes experienced by the non-rotating models of the two systems.

After the H-rich skin has been completely transferred to the WD, a He-rich buffer is piled up onto the WD. We find that, when the WD total mass becomes 0.8934\,\msun, a He-flash is ignited at the mass coordinate $M_{ig}=0.794$\,\msun, where the density is $\rho_{ig}=1.15\times 10^6$\,\gcc.
Comparing these values with those derived for the non-rotating model of \ptf~ in Paper I (\mwd=0.888\,\msun, $\Delta M_{He}=0.137$\,\msun, $M_{ig}=0.8$\,\msun\ and $\rho_{ig}=8\times 10^5$\gcc) it is evident that a helium detonation will develop. The value we got for the He-rich buffer at the epoch of He-ignition\footnote{Consistently with Paper I, we assume that H or He are ignited when the energy released per unit time by CNO cycle or 3$\alpha$-reactions, respectively, first exceeds 100 times the surface luminosity (see Sect. 2 in Paper I).} is consistent with the results reported in \citet{hillman2026} (see their Tab. 3).
As in Paper I, we follow the evolution of the accreting white dwarf after He-ignition with the hydrodynamical 1D code TSUHN \citep{bravo2019}.
About 22.6 s after the temperature in the hot helium layer reached $2\times10^9$ K, helium detonation has propagated outwards by 0.05 M$_\sun$ through the helium cap, and the overpressure induces an inward helium detonation. 
Converging shock waves are strong enough to ignite carbon detonation inside the white dwarf. 
Finally, the star is destroyed, leading to a peculiar Type Ia supernova-like event, with a kinetic energy of $0.98\times 10^{51}$ erg and an ejected mass of $^{56}$Ni amounting to 0.243 M$_\sun$. 
The composition of the ejecta is rich in intermediate-mass elements: 
silicon is overproduced by a factor of 1.8 relative to iron, 
while $^{44}$Ca, $^{48}$Ti, $^{74}$Se, $^{78}$Kr, and $^{80}$Kr are all overproduced more than tenfold with respect to $^{56}$Fe 
(all features as compared to the standard solar abundances).

Our non-rotating model differs from most models found in the literature, mainly because the initial model WD has a thin helium skin and the accretion rate isn't constant, but it is determined by the evolution and increases with time. Nevertheless, we find that our model reasonably agrees with the 1D model 7C of \citet{woosley_kasen_dd11} at the He-detonation epoch. The overproduction factors with respect to iron for \isotope{44}{Ca} and \isotope{48}{Ti} are virtually similar, while for \isotope{28}{Si},  \isotope{78}{Se}, \isotope{78}{Kr}, and \isotope{80}{Kr} they are higher.
Probably this may be attributed to the difference of nuclear networks adopted in the computations.

At the explosion epoch, the donor mass is 0.315\,\msun. 
If the donor survives the explosion, it is expected to move away from the original location of the progenitor binary system with a velocity equal to its orbital velocity around the center of mass of the system. 
Eventually, the velocity of the former donor can increase via the momentum transferred to the star by the ejecta of the exploding companion. 
However, the kick produced by the explosion is expected to affect the orbital velocity by no more than 10-20\% \citep[see for instance][]{liu2013,bauer2019,wong2025}.

We found that, by neglecting the latter contribution \citep[see for instance][]{neunteufel2017}, the donor will become a hypervelocity star with $v_{sdB}$=747\,${\rm km~s^{-1}}$.

\section{Rotating model}\label{s:rot}

\begin{figure}[b]
	\centering
	\includegraphics[width=\columnwidth]{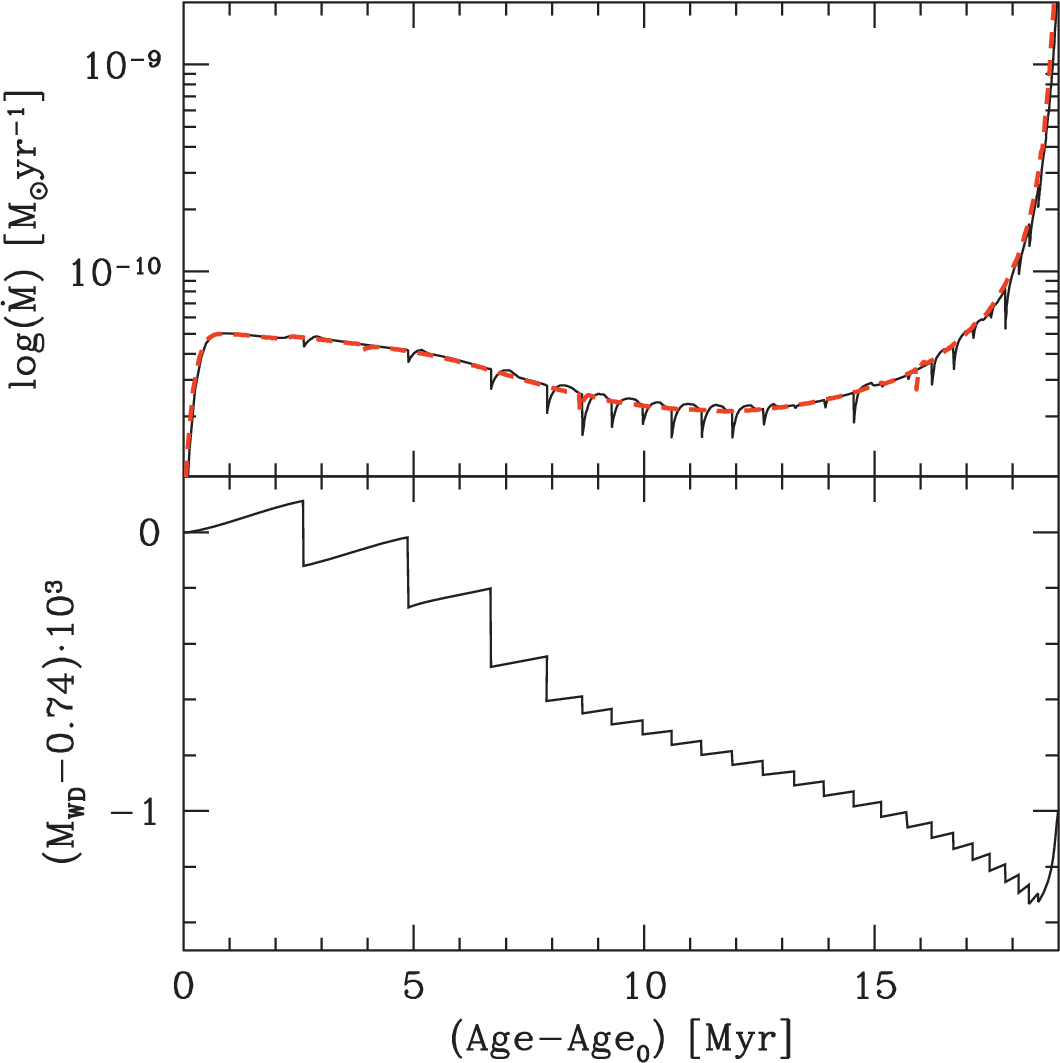}
	\caption{Evolution of the rotating CO WD during the H-accretion phase. From top to bottom we report the mass transfer rate from the donor (in \msyrm) and the variation of the WD total mass defined as $M_{WD}^*=10^{3}\times\left(\frac{M_{WD}}{M_\odot}-0.74\right)$. Zero-age ($\mathrm{Age_0}$) corresponds to the epoch of RLOF by the sdB star. For the sake of comparison, we report in the upper panel the mass transfer rate for the non-rotating model (red dashed line).}
	\label{fig2}
\end{figure}

We recomputed the evolution of the CD-30 system accounting for the effects of rotation on the physical and chemical properties of the accretor as outlined in Paper I (henceforth, ``rotating model'').
As a first step, we account for rotation-driven instabilities only, namely, Eddington-Sweet (ES), Goldreich-Shubert-Fricke (GSF), Solberg-H{\o}iland (SH), and shear (both secular and dynamical, SS and DS, respectively) instabilities.
As discussed in Paper I (see Sect. 5.1), when the accreted matter deposits angular momentum onto the accretor, the local gradient of the angular velocity drives the onset of shear instabilities, which determines not only the inward transport of angular momentum but also the mixing of the accreted matter with the outermost layers of the WD.
During the H-accretion phase, this determines, with respect to the non-rotating model above, the growth of the H-rich zone so that the H-flash is ignited when a smaller amount of matter has been accreted, even if the ignition density is higher and, hence, the energy released is larger. 
During the subsequent RLOF episode all the accreted layers as well as the most external zone of the accretor are ejected from the system, so that the WD mass is secularly reduced.
We find that, soon after the sdB star fills its Roche lobe, the mass transfer rate rapidly increases to \mdot$\sim 5\times 10^{-11}$\msyrm (see upper panel in Fig.~\ref{fig2}). During the H-accretion phase, the accretor in the CD-30 system experiences 24 H-flashes followed by RLOF episodes during which the WD total mass reduces to 0.7387\,\msun\, (see lower panel in Fig.~\ref{fig2}).
\begin{table}
	\caption{Same quantities as in Table~\ref{hflnor}, but for  selected H-flash episodes of the rotating model.}
	\label{hflrot}      
	\centering                   
	\begin{tabular}{l c c c c }
		\hline\hline                 
		\# Fl.            &    1   &   5   &   16  &   23  \\ 
		\hline
		$\Delta T$        &   2.60 &  0.77 &  0.57 &  0.23 \\
		$\rho_{ig}$       &   4.53 &  2.00 &  1.93 &  2.49 \\
		$T_H^{max}$       &   1.66 &  1.33 &  1.36 &  1.49 \\
		$L_H^{max}$       & 388.97 & 29.82 & 26.80 & 46.71 \\
		$\Delta M_{acc}$  &   0.63 &  0.21 &  0.17 &  0.30 \\
		$\Delta M_{lost}$ &   1.84 &  0.61 &  0.55 &  0.68 \\
		<\mdote>          &   2.42 &  2.68 &  2.96 & 13.04 \\
		$\eta$            &  -1.92 & -1.97 & -2.30 & -1.25 \\		
		\hline                 
	\end{tabular}
\end{table}

After that, the WD experiences one additional H-flash, but, due to the low H-abundance in the accreted matter (mass fraction X(H) < $10^{-3}$), the delivered energy is too low to produce a large expansion of the external layers, so that the matter is not ejected from the system.
In Table~\ref{hflrot}, we report some relevant properties of the system during four selected H-flashes followed by RLOF\footnote{A complete table listing the same properties for all the 24 H-flashes is reported in Appendix~\ref{a:tables}.}.
By inspecting this table and Fig.~\ref{fig2}, it comes out that the physical properties of the accreting WD are very variable. 
This is due to the combination of mass transfer rate variations, H-abundance in the accreted matter, and efficiency of secular shear instabilities, as discussed in Sect. 5.1 of Paper I.

\begin{table}[t]
	\caption{Selected properties of the accreting rotating WD during evolution throughout six recurrent He-flashes.}
	\label{heflashes}      
	\centering                   
	\begin{tabular}{l c c c c c c }
		\hline\hline                 
		\# Fl.                &   1   &   2   &   3   &   4   &   5   &   6   \\ 
		\hline
		\mwd                  & 0.739 & 0.724 & 0.708 & 0.703 & 0.694 & 0.689 \\
		$\Delta T$            &  1.33 &  2.52 &  3.65 &  1.77 &  1.00 &  0.41 \\
		Per.                  & 27.00 & 25.65 & 23.20 & 16.72 & 12.14 &  8.63 \\
		$\Delta M_{acc}$      &  1.16 &  6.24 &  8.24 &  5.10 &  5.08 &  3.78 \\
		$\Delta M_{lost}$     &  2.60 &  7.85 &  8.78 &  5.98 &  5.55 &  3.77 \\
		$\Delta M_{He}^{pk}$  &  2.67 &  7.92 & 10.02 &  6.66 &  6.16 &  4.41 \\
		<\mdote>              &  0.87 &  2.47 &  2.26 &  2.88 &  5.07 & 9.20 \\
		$\rho^{ig}_{He}$      &  1.06 &  2.71 &  2.96 &  1.79 &  1.49 & 1.05  \\ 
		$T_{He}^{max}$        &  3.78 &  3.56 &  3.01 &  3.33 &  3.20 & 4.46  \\
		$L_{He}^{max}$        &  3.88 & 10.22 & 28.58 &  2.30 &  2.92 & 0.17  \\
		$\eta$                & -1.24 & -0.26 & -0.07 & -0.17 & -0.09 & 0.00  \\
		\hline                 
	\end{tabular}
	\tablefoot{For each flash episode, we report the mass of the WD at the beginning of the cycle (\mwd\ in \msun), the duration of the accretion phase ($\Delta T$ in Myr), the orbital period at the reonset of mass accretion (Per., in min.), the mass accreted between two successive flashes ($\Delta M_{acc}$ in $10^{-2}$\,\msun), the mass lost during the flash-driven RLOF ($\Delta M_{lost}$ in $10^{-2}$\,\msun), the mass of the He-rich zone with X(\isotope{4}{He}) > 0.1 ($\Delta M_{He}^{pk}$ in $10^{-2}$\,\msun), the average accretion rate before the onset of RLOF (<\mdote> in $10^{-8}$\msyrm), the ignition density for He-burning ($\rho^{ig}_{He}$ in $10^5$\,\gcc), the maximum temperature ($T_{He}^{max}$ in $10^{8}$ K) in the burning shell, the maximum energy per unit time delivered via He-burning ($L_{He}^{max}$ in $10^{12}$\,\lsun) during the He-flash, and the retention efficiency $\eta$.}
\end{table}

After the H-rich layer of the sdB star has been completely lost, He-rich matter is accreted.
\begin{figure}[b]
	\centering
	\includegraphics[width=\columnwidth]{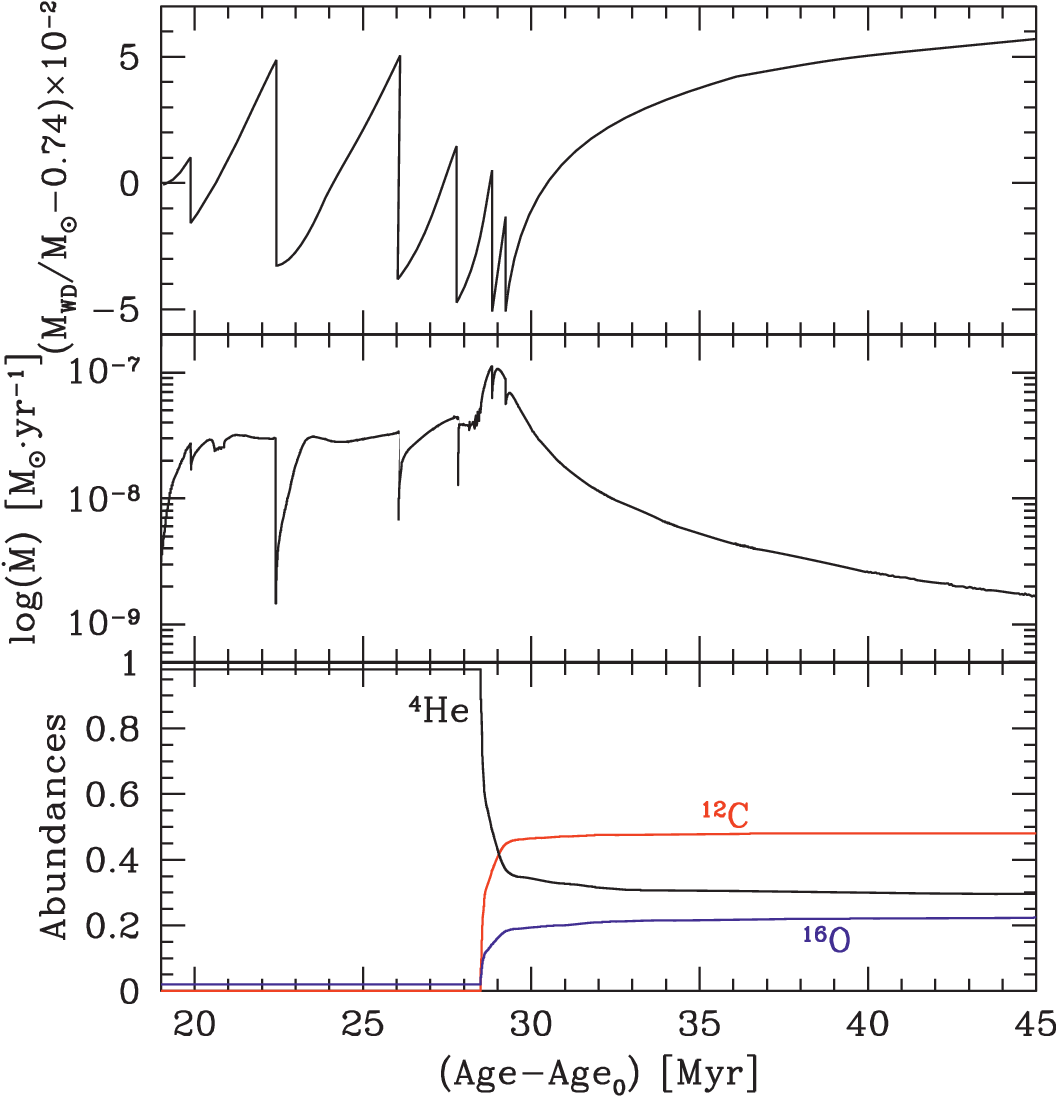}
	\caption{Time dependence of the WD total mass (upper panel), of the mass transfer rate from the companion (middle panel), and of the chemical composition of the accreted matter (lower panel) during the He-rich mass accretion phase in the rotating model.}
	\label{fig3}
\end{figure}
In this case, due to rotation-induced instabilities, the accreted matter is mixed with the external layers of the WD so that, at variance with the non-rotating model above, a He-flash is ignited when a small amount of mass has been transferred $\Delta M_{acc}$=0.012\msun, to be compared with $\Delta M_{acc}$=0.153\msun\, for the non-rotating model.
In fact, as discussed in Paper I (see Sect. 5.2), the zone involved in the He-flash is definitively larger than the accreted layer, as it is evident by comparing the accreted mass and the mass of the zone where the helium abundance by mass fraction is larger than 0.1 ($\Delta M_{acc}$ and $\Delta M_{He}^{pk}$ in Table~\ref{heflashes}, respectively). 
Moreover, in the rotating model, the ignition density is lower due to rotation and the dissipation of rotational energy heats up efficiently the He-shell. 
As a result, the flash does not turn into a detonation, but it is very strong and triggers a RLOF during which all the accreted matter as well as the most external layers of the WD are ejected from the system. 
During the He-rich accretion phase, the accretor in the CD-30 system experiences six He-flashes followed by RLOF episodes (see the upper panel in Fig.~\ref{fig3}).
In Table~\ref{heflashes}, we report several physical properties of the accreting WD during the He-rich accretion phase.
After six He-flashes, the WD total mass has been reduced to 0.689\,\msun\, and the mass transfer rate at that epoch is $1.066\times 10^{-7}$\myr.
During the subsequent evolution, He-rich matter continues to be accreted, but the WD never attains the physical conditions suitable for igniting He again due to both the continuous decrease of the mass transfer rate (see middle panel in Fig.~\ref{fig3}) and the reduction of the He abundance in the accreted matter (see lower panel in Fig.~\ref{fig3}). 

We observe that none of the AM CVn systems studied so far exhibits a clear carbon signature \citep[see, for example. ][]{nelemans2010, green2025}. 
This could indicate that the He-WD channel likely dominates over the He-star channel as a formation scenario for AM CVn systems. However, we also note that in the He-star channel, C dominates in transferred matter only if the donor was significantly evolved at the RLOF. In this case, the transfer of C-enriched material would occur when the luminosity of both components of AM CVn stars is so low that spectroscopic analyses may be challenging.

We stopped the computation when the mass of the donor had been reduced to $M_{don}=0.0197$\,\msun, as the adopted input physics became inadequate to describe extremely cool objects. 
At this epoch, the mass accretion rate is \mdot=$1.402\times  10^{-11}$\,\msyrm\, and the WD mass is 0.842\,\msun. 
\begin{figure}[t]
	\centering
	\includegraphics[width=\columnwidth]{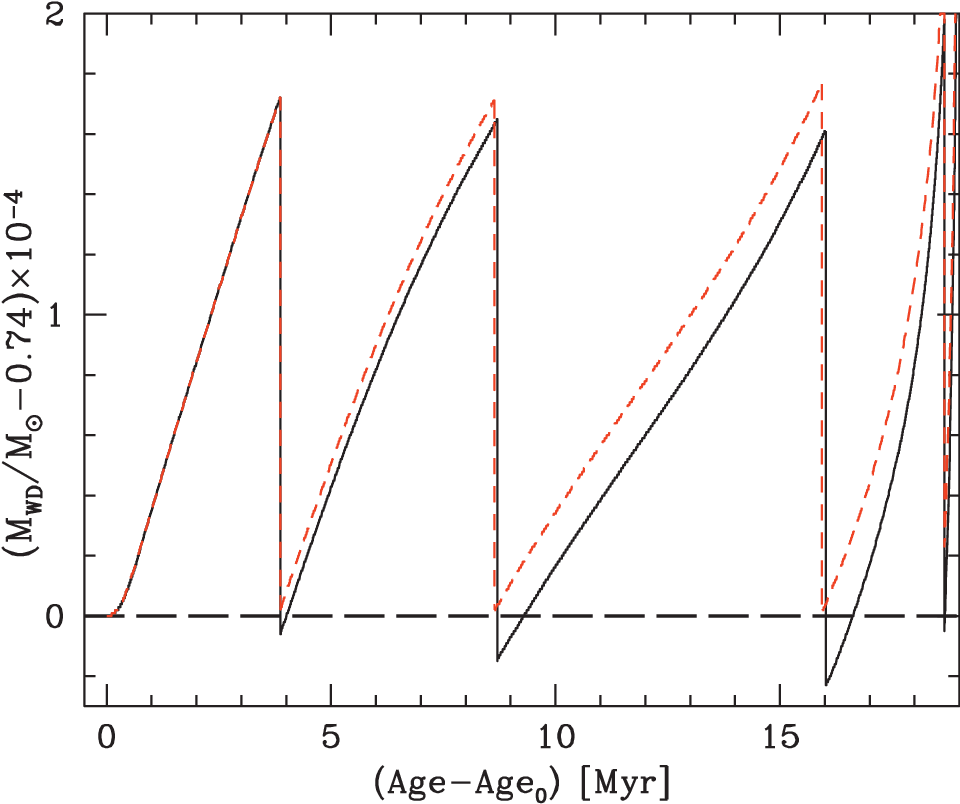}
	\caption{Time evolution of the WD total mass during the H-rich mass accretion phase for the non-rotating model (red dashed line) and the magnetic model (solid black line).
	}
	\label{fig4}
\end{figure}

\section{Magnetic model}\label{s:ts}

According to \citet{spruit1999,spruit2002} \citep[see also ][for independent confirmations]{hartogh2020,petit2023,petit2024}, the presence of strong differential rotation inside radiative regions of a star can generate small magnetic field seeds that can be amplified into a toroidal magnetic field via a dynamo process by an instability of the field itself, most probably the Tayler instability \citep{tayler1973}.
The torque associated with this field is responsible for the transport of angular momentum through stellar interiors. Such a mechanism, usually addressed as Tayler-Spruit mechanism, sets in only if the local angular velocity gradient overcomes the stabilizing effects of buoyancy forces and the molecular weight gradients. 
Such a condition is mathematically expressed by requiring that the dimensionless differential rotation rate, defined as $q=\frac{\partial\ln\omega}{\partial\ln r}$\footnote{Here, $\omega$ represents the angular velocity and $r$ is the radial coordinate.}, is larger than a critical value that accounts for both the chemical and thermal stratification inside a star \citep[see the discussion in][]{spruit2002}.
{The Tayler-Spruit dynamo is responsible for the transport of angular momentum via magnetic stresses, while macroscopic motion of stellar material remains largely inefficient \citep[see the discussion in Sect. 3.2 of][]{spruit2002}.

As matter and angular momentum are deposited onto the accreting WD, a local rotation rate gradient forms. Under these conditions, according to order-of-magnitude estimates \citep{spruit2002} and detailed stellar evolution models \citep{neunteufel2017}, the Tayler-Spruit mechanism sets in well before the physical conditions suitable for efficient shear instabilities are attained.
Moreover, magnetic instabilities operate on a very short timescale, redistributing angular momentum and, thus, enforcing rigid rotation along the entire WD. 
Hence, as discussed in Paper I, DS-, SS-, GSF-, and SH-instabilities are suppressed, while ES circulation remains active, even if its strength in the accreted layers is largely reduced due to the decrease of the local value of the angular velocity. In addition, also the shear factor, $\sigma=\frac{\partial\omega}{\partial\ln r}$, is largely reduced, so that local release of thermal energy due to the dissipation of rotational energy is practically zeroed.
Finally, \citet{spruit2002} remarks that magnetic stresses, responsible for the angular momentum transport, are definitively more efficient than the stresses related to the motion of matter, induced by the dynamo process, responsible for the mixing of chemical species. 
This implies that the efficiency of mixing is significantly reduced. 
Hence, at variance with the rotating model above, the evolution of the accreting WD is affected mainly by the reduction of the effective gravity due to rotation (lifting effect of rotation). 
\begin{table}
	\caption{Selected properties of the accreting WD during the evolution throughout recurrent H-flashes of the magnetic model.}
	\label{hflashes}      
	\centering                   
	\begin{tabular}{l c c c c }
		\hline\hline                 
		\# Fl.                &   1   &   2   &   3   &   4   \\ 
		\hline
		$\Delta M_{acc}$      & 1.72  & 1.71  & 1.76  & 2.21  \\
		$\Delta M_{lost}$     & 1.78  & 1.80  & 1.84  & 2.03  \\
		$\omega_{ig}$         & 1.72  & 2.00  & 2.31  & 2.72  \\
		$\rho_{ig}$           & 2.21  & 2.26  & 2.38  & 3.55  \\
		$T_{H}^{max}$         & 1.45  & 1.46  & 1.52  & 1.77  \\
		$\Delta M_{H}^{pk}$   & 1.81  & 1.85  & 1.88  & 2.31  \\
		$\eta$                & -0.03 & -0.05 & -0.05 & 0.08  \\
		\hline                 
	\end{tabular}
	\tablefoot{For each flash episode, we report the mass accreted between two successive flashes ($\Delta M_{acc}$ in $10^{-4}$\,\msun), the mass lost during the flash-driven RLOF ($\Delta M_{lost}$ in $10^{-4}$\,\msun), the angular velocity ($\omega_{ig}$ in $10^{-3} {\rm rad\cdot s^{-1}}$) and the density ($\rho_{ig}$ in $10^{3}$\,\gcc) at the H-ignition point, the maximum temperature ($T_{H}^{max}$ in $10^{8}$ K) attained during the H-flash, the mass of the H-rich zone with X(H) > 0.1 ($\Delta M_{H}^{pk}$ in $10^{-4}$\,\msun), and the retention efficiency $\eta$.}
\end{table}

\begin{figure*}[t]
	\sidecaption
	\includegraphics[width=12cm]{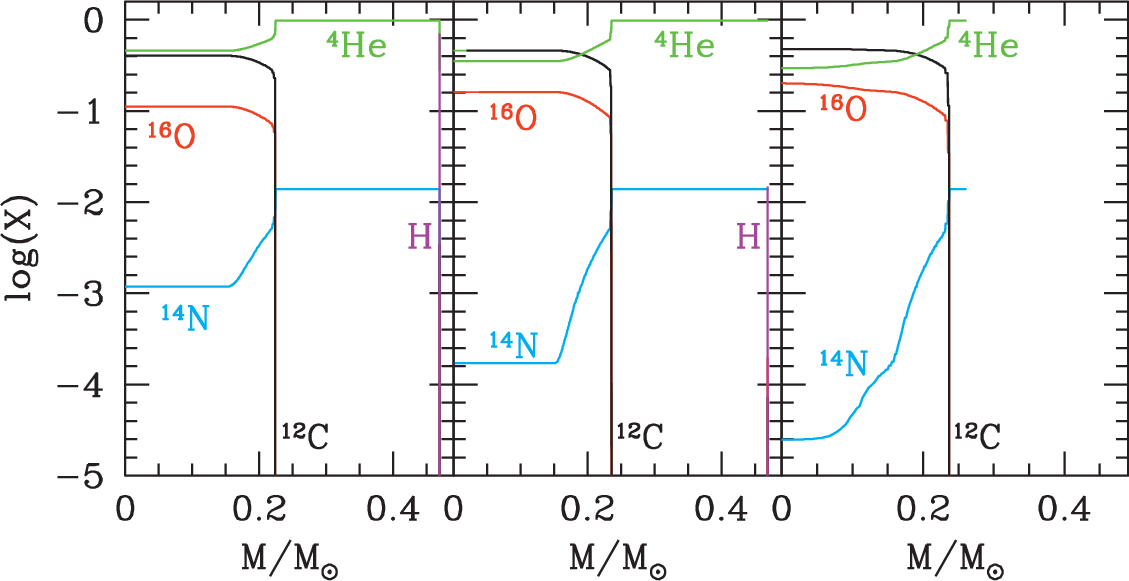}
	\caption{Mass fraction abundance of selected isotopes along the sdB donor in CD-30 at three epochs during the evolution of the magnetic model: 
	onset of the very first RLOF episode (left panel);
	start of the transfer of H-deprived material (middle panel);
	onset of the last mass transfer episode driving a He-flash (right panel).}
	\label{fig5}
\end{figure*}

In Fig.~\ref{fig4}, we compare the variation of the WD total mass during the H-rich mass accretion phase for the non-rotating model (dashed red line) and a rotating model including magnetic instabilities (hereinafter ``magnetic model'').
As can be noticed, during the first cycle, a H-flash is ignited in both models when virtually the same amount of H-rich material has been accreted.
Since the accreted mass is small, in the magnetic model the amount of angular momentum acquired by the accretor is small, so that, due to efficient angular momentum transport related to the Tayler-Spruit mechanism, the angular velocity in the H-rich accreted layer is very low and, hence, the lifting effects of rotation are almost negligible. Further inspection of Fig.~\ref{fig4} and Table~\ref{hflashes} reveals that the RLOF episode occurring in the magnetic model results in a small erosion of the initial WD. This is due to the partial mixing produced by magnetic instabilities at the border of the accreted layer and the original WD, which determines a small increase of the mass of the H-rich zone compared to the amount of accreted matter (see $\Delta M_{H}^{pk}$ in Table~\ref{hflashes}).

During the following two accretion cycles, the angular velocity in the accreted layers slowly increases due to the continuous deposition of angular momentum, and, hence, the amount of accreted matter slightly increases and the resulting H-flash delivers a larger amount of energy (the maximum attained temperature in the H-rich zone increases), so that the retention efficiency becomes negative. 

Finally, during the fourth H-flash episode, the mass fraction abundance of hydrogen in the accreted matter decreases, as the H-rich buffer on the top of the sdB companion has been almost completely eroded. 
As a consequence, a larger amount of mass has to be accreted to trigger a new H-flash, which is ignited at a definitively larger density.
Notwithstanding, the amount of energy per unit mass injected into the accreted layers is smaller as compared to the third H-flash, so that the mass lost during the subsequent RLOF episode is smaller than the accreted one.
Like the non-rotating model, the magnetic one experiences an additional H-flash, which delivers a very small amount of energy, so that an additional RLOF episode does not occur. 

After the H-rich buffer on the sdB star has been completely transferred, accretion of He-rich matter onto the WD begins.
As discussed in \cite{yung2008}, when He-rich matter starts to be transferred to the accretor, the mass transfer rate rapidly increases, determining the decrease of the surface radius and luminosity of the donor. Due to mass transfer, the nuclear burning efficiency in the center of the sdB star rapidly decreases \citep[see also the discussion in ][]{brooks2015}. Thus, the chemical composition of the donor remains practically unaltered during the evolution following the new RLOF episode, as displayed in Fig.~\ref{fig5}, where we report the abundance profile for selected isotopes in the sdB donor at three different epochs during the evolution of the magnetic model for the CD-30 system. This plot, 
clearly shows the effect of induced overshooting and consequent semiconvection operating at the border of the convective core \citep{castellani1971}, responsible for the smooth transition of the profiles at the mass coordinate $\sim$0.2\,\msun. As discussed in \cite{yung2008} and \cite{brooks2015}, the abundance of the various isotopes depends mainly on the epoch at which the He-core starts to be eroded via mass transfer, which is determined mainly by the initial orbital period of the considered binary.

\begin{figure}[t]
	\centering
	\includegraphics[width=\columnwidth]{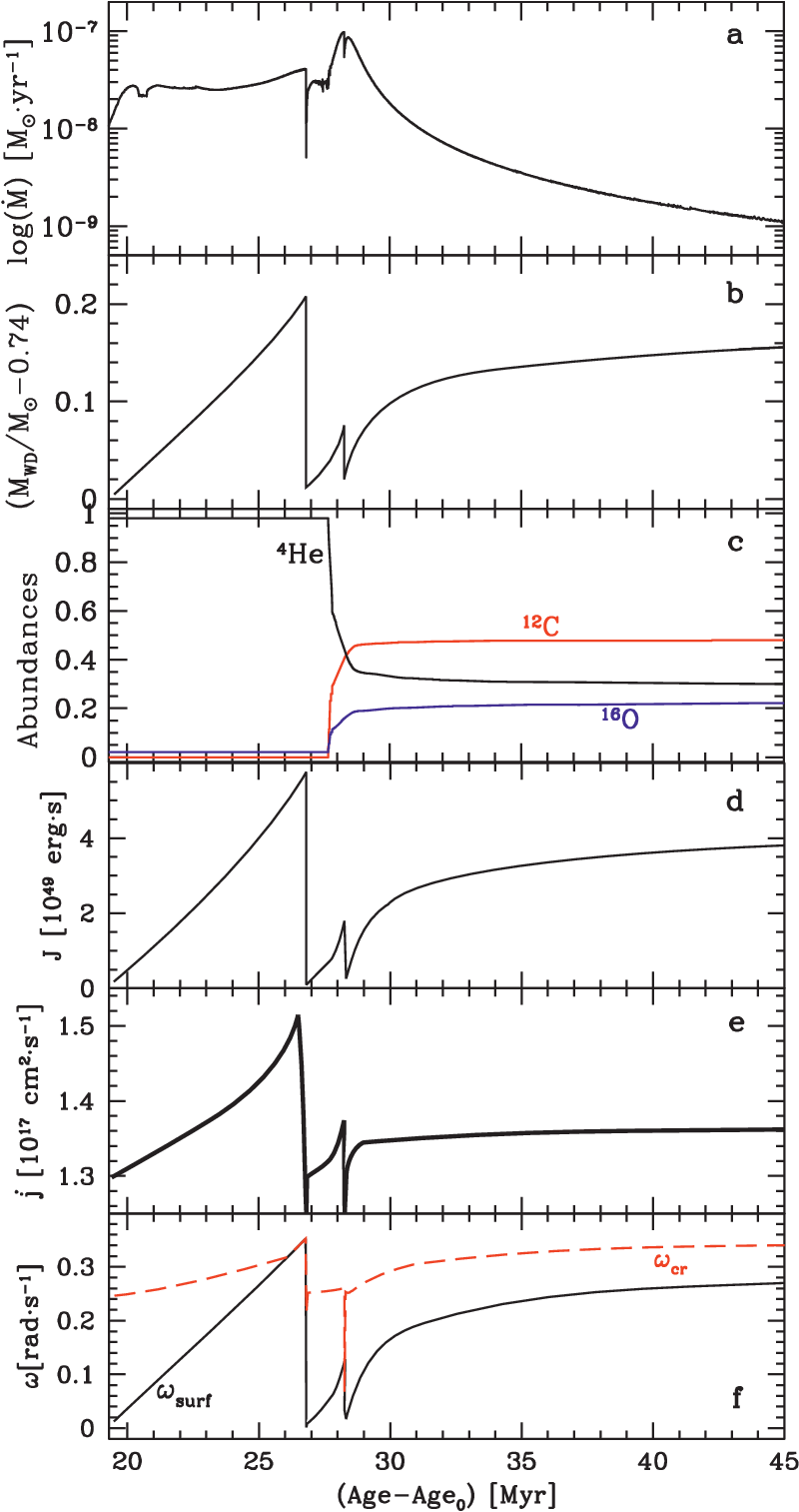}
	\caption{Time evolution of selected physical properties of the magnetic model during the He-accretion phase. From top to bottom, we plot: the mass transfer rate (in \msyrm~ -- panel a), the total mass of the accreting WD (in \msun~ -- panel b), the chemical composition of the accreted matter (panel c), the total angular momentum of the accreting WD (in $10^{49} {\rm erg\cdot s}$ -- panel d), the specific angular momentum of the accreted matter (in $10^{17} {\rm cm^2\cdot s^{-1}}$ -- panel e), and the angular velocity (black line) and the critical rotation rate (red line) at the surface (in ${\rm rad\cdot s^{-1}}$ -- panel f).}
	\label{fig6}
\end{figure}
As displayed in panel (a) of Fig.~\ref{fig6}, after He-rich matter begins to be accreted, the mass transfer rate rapidly increases to an almost constant value \mdot$\sim 2\times 10^{-8}$\,\msyrm.
The accreted matter deposits angular momentum so that the WD total angular momentum increases (see panels (d) and (e) in Fig.~\ref{fig6}).
Owing to the very efficient angular momentum transport via the Tayler-Spruit mechanism, the angular velocity rapidly increases in the entire accreting WD (see the lower panel in Fig.~\ref{fig7}).
Due to the lifting effects of rotation in the accreted layer, local compressional heating is reduced, so that the mass of the He-rich buffer continues to increase well beyond the value attained in the non-rotating model (see panel (b) in Fig.~\ref{fig6}).
When the total mass of the WD attains 0.895\,\msun, the angular velocity at the surface of the WD reaches its critical value, $\omega_{cr}=\sqrt{\rm{GM_{WD}/R_{WD}^3}}$, where \mwd~ and \rwd~ are the total mass and the surface radius of the accreting WD, respectively (see panel (f) in Fig.~\ref{fig6}). 
Hence, during further evolution the amount of angular momentum gained per unit time is smaller (see the decrease of $\dot{J}$ at ${\rm(Age-Age_0)}\simeq$ 26.5 Myr -- panel (e) in Fig.~\ref{fig6}).
At the same time, the mass transfer rate slightly increases to $\sim 4\times 10^{-8}$\,\msyrm\ (see panel (a) in Fig.~\ref{fig6}).
The interplay of the reduced spin-up per unit time and the increase of the accretion rate results in a more efficient compression in the most external layers of the He-rich accreted buffer, so that, when the WD attains a total mass of \mwd=0.9476\,\msun, He-burning is ignited via a very strong flash at the mass coordinate 0.8407\,\msun, in the middle of the He-rich accreted layer (see the upper panel in Fig.~\ref{fig7}).
At the ignition epoch, the density at the He-ignition layer is $\rho^{ig}_{He}=6.99\times 10^5$\,\gcc, a value definitively larger than that obtained for He-flashes in the rotating model (see Table~\ref{heflashes}), but not large enough to trigger a detonation.
\begin{figure}[t]
	\centering
	\includegraphics[width=\columnwidth]{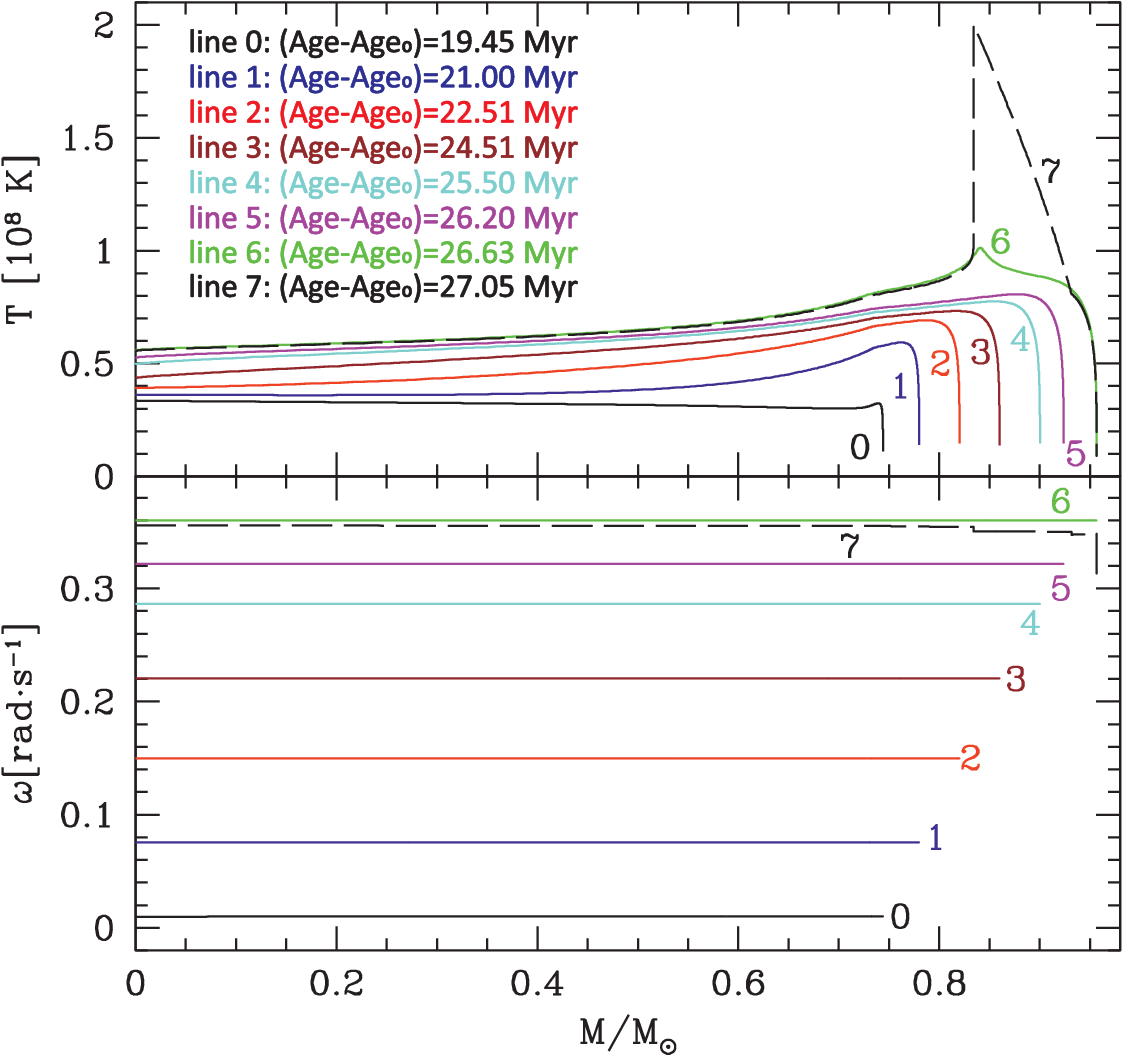}
	\caption{Temperature (upper panel) and angular velocity (lower panel) profiles for selected models during the first He-rich matter accretion episode in the magnetic model. Lines labeled 5 correspond to the epoch when the accretor attains the critical value of angular velocity at the surface; lines labeled 6 correspond to the epoch when flash-driven convection sets in. In the upper panel, we report for each model the time elapsed from the onset of mass transfer ($\mathrm{Age-Age_0}$ in Myr).}
	\label{fig7}
\end{figure}

In order to verify the latter statement, we implemented the hydrodynamic equation as described in \cite{mezzacappa1993} in a 1D explicit Lagrangian code. We allow for the effects of rotation under the assumption of local conservation of angular momentum (no angular momentum transport accounted for).
We make use of a nuclear network derived from the 19-isotopes network described in \citet{weaver1978}. In particular we explicitly account for both the $(\alpha,\gamma)$ and $(\alpha,p)(p,\gamma)$ reactions, relaxing the assumption of steady-state flow of protons through intermediate isotopes; in addition we explicitly account for all the isotopes involved in the CNO cycle and in the pp-chain. We also include the \isotope{13}{N}${\textrm(\alpha,p)}$\isotope{16}O, as it has been suggested that such a reaction could boost significantly the energy production when the temperature during the He-flash increases above $10^9$ K \citep[see][]{woosley_kasen_dd11,shen2014b}.
We find that the maximum attained temperature during the He-flash is 
$T_{He}^{max}=1.36\times 10^9$ K. We also find that no strong shocks develop so that the flash does not turn into a detonation, while the burning front propagates outward conductively, consuming several $10^{-3}$\,\msun\, of He-rich material before the structure starts to expand. 
The effect of including the \isotope{13}{N}${\textrm(\alpha,p)}$\isotope{16}O reaction is to increase the maximum attained temperature during the flash by $\sim 8\times 10^6$ K. We note that the ignition density for the He-flash in our simulation is in agreement with that estimated by \citet{woosley_kasen_dd11} for the occurrence of a He-nova or a He-deflagration.
We repeat the computation above by adopting a larger nuclear network including \isotope{12-14}{C}, \isotope{13-15}{N}, \isotope{15-18}{O}, \isotope{18-19}{F},\isotope{20-22}{Ne}, \isotope{21-23}{Na}, \isotope{24-26}{Mg}, \isotope{25-27}{Al}, \isotope{27-30}{Si}, \isotope{31}{P}, \isotope{32}{S}, \isotope{35}{Cl}, \isotope{36}{Ar}, \isotope{39}{K}, \isotope{40}{Ca}, \isotope{43}{Sc}, \isotope{44}{Ti}, \isotope{47}{V}, \isotope{48}{Cr}, \isotope{51}{Mn}, \isotope{52-56}{Fe}, \isotope{55-56}{Co}, \isotope{56}{Ni}, linked by all the relevant $\alpha-,\ p-$ and ${n-}$capture and weak processes. We find that also in this case the flash does not turn into a detonation, the maximum attained temperature modestly increasing to $T_{He}^{max}=1.40\times 10^9$ K.
After the epoch of maximum temperature, timesteps are lower than $10^{-3}$ s, as they are determined by the Courant-Friedrichs-Lewy convergence condition \citep{courant1928}. As a consequence, our explicit code is inadequate to follow the further evolution of the accretor, which is expected to last $\sim$ 40 yr. For this reason, as the He-flash does not become dynamical, we continue the computation of this model with the FuNS code.
As the heating timescale becomes comparable to the turnover timescale, the contribution of convection to the total transport of thermal energy freezes out. In order to account for such an occurrence we make use of the approach described in \cite{wood1974}.

We find that the maximum temperature attained during the first He-flash is $T_{He}^{max}=1.21\times 10^9$ K (slightly smaller than that obtained with the explicit hydrodynamic code), while the maximum energy per unit time delivered via He-burning is $L_{He}^{max}=8.39\times 10^{14}$ \lsun.
Due to the large amount of energy deposited above the He-burning shell, the WD expands very rapidly, filling its Roche lobe.
During this phase, due to the very short evolutionary timescale, the mass coordinate of the He-burning shell remains fixed, while the buffer above the shell is reduced at a very high rate.
When the WD total mass attains \mwd$\simeq$ 0.848\,\msun, the evolutionary timescale becomes longer and the He-burning shell can propagate inward, thus continuing to inject nuclear energy so that mass ejection via RLOF continues, even if at a lower rate. 
The RLOF episode ends when the total mass of the WD has been reduced to \mwd=0.7517\,\msun. 
The resulting accumulation efficiency is $\eta=3.4$\%.
During the RLOF episode, the layers above the burning shell expand and, hence, the local angular velocity decreases, but the very efficient angular momentum transport via magnetic instability enforces once again rigid rotation. 
During this mass loss phase, angular momentum is extracted from the core and redistributed in the He-rich envelope so that at the end of the RLOF episode the average angular velocity along the entire WD is $\simeq 0.08\ {\rm rad\cdot s^{-1}}$ (see panels (d) and (f) in Fig.~\ref{fig6}).

During the following evolution, mass transfer from the sdB companion resumes and, after accreting $\Delta M\simeq 0.064$\,\msun, the WD experiences a new He-flash that is ignited at a mass coordinate 0.787\,\msun, where the density is $\rho^{ig}_{He}=2.23\times 10^5$\,\gcc.
Even if this second flash is much weaker than the first one ($L_{He}^{max}=4.6\times 10^{11}$ \lsun , $T_{He}^{max}=5.35\times 10^{8}$ K), it causes a new RLOF episode, during which about 0.056\,\msun\ of He-rich material is ejected from the system. The corresponding retention efficiency is $\eta=13.13$\%.

The second He-flash episode is very different from the previous one for several reasons.
First, the thermal content of the accreting WD at the reonset of the mass transfer process is completely different because the first flash episode has injected nuclear energy in the external layers of the accreting WD, which exhibits now a local maximum in the temperature profile.
Second, the mass transfer rate during this accretion episode rapidly increases to $9.78\times 10^{-8}$\,\msyrm\, (see panel (a) in Fig.~\ref{fig6}), so that the local compressional heating occurs more rapidly, overcoming the lifting effects of rotation.
At the end, during the mass transfer, zones of the sdB donor previously involved in the convective central He-burning are lost so that the accreted matter is carbon- and oxygen-rich and He-poor (see panel (c) in Fig.~\ref{fig6}).

After the end of the second RLOF episode, mass transfer resumes, but at a continuously decreasing rate, as displayed in panel (a) of Fig.~\ref{fig6}.
In addition, the He-abundance in the accreted matter is reduced (see panel (c) in Fig.~\ref{fig6} and the right panel in Fig.~\ref{fig5}), so that the WD never attains the physical conditions suitable for an additional He-flash. The fate of the WD in CD-30 is the formation of a CO core, capped by a massive He/C/O buffer of about 0.194\,\msun\ (see Fig.~\ref{fig8}).
Like for the rotating model in Section \ref{s:rot}, due to the inadequacy of the adopted input physics, we were forced to stop the computations when the donor mass decreased to $M_{\rm don}=0.0197$\,\msun. 
\begin{figure}[t]
	\centering
	\includegraphics[width=\columnwidth]{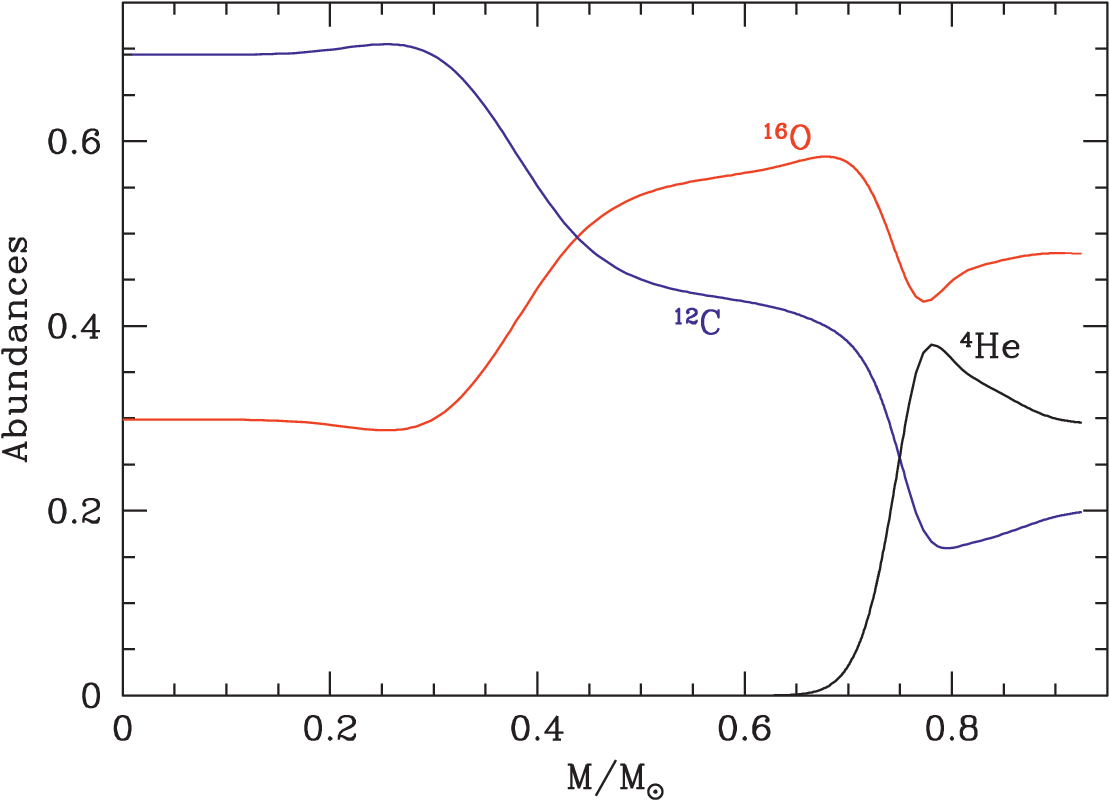}
	\caption{Mass fraction abundance of \isotope{4}{He} (black line), \isotope{12}{C} (blue line) and \isotope{16}{O} along the last computed structure of the magnetic model.}
	\label{fig8}
\end{figure}

\section{Discussion and conclusion}\label{s:concl}

We have investigated the future evolution of the sdB+WD binary \cdth.
The present work is an extension of the paper by \citet{piersanti2024}, where we studied the similar system \ptf.
In addition to consideration of the effects of rotation upon physical and chemical properties of the accretor, we studied a model that accounts for magnetic instabilities and their effect upon the angular velocity profile of the accreting WD.
In the rotating non-magnetic models, either magnetic or non-magnetic, it becomes impossible for the WD to experience a detonation in the layer of accreted helium leading to an explosive event of a SN~Ia scale. 
We find that the mass transfer from the sdB companion results in the formation of a WD with a CO core surrounded by a massive He/C/O buffer.
This outcome does not depend on the physical mechanisms (rotation-driven 
and/or magnetically-driven instabilities) defining the transport of angular momentum, even if the physical properties of the accretor, as well as its evolution are different. 
In particular, we find that, in the rotating model, the original WD is secularly eroded during recurrent H- and He-flashes, while the mass of the magnetic model remains practically unchanged during accretion of H-rich matter and slightly increases during the He-accretion phase.
In fact, in the magnetic model, the Tayler-Spruit instability enforces rigid rotation, so that the mixing of the accreted matter with the underlying core is largely suppressed and, hence, each flash episode occurs well inside the accreted layer.
On the other hand, as the shear factor in the magnetic model is practically absent, heating via dissipation of rotational energy does not occur.
This implies that the physical properties during each mass transfer episode in the magnetic model depend only on the interplay between the mass transfer rate, which determines the compressional heating rate due to accretion, and the lifting effect of rotation, which defines the compressibility of the accretor. 
For this reason, the magnetic model experiences two He-flashes only, the first one after accreting a very massive He-buffer, typical for a He-detonation in a non-rotating structure. 

The efficiency of the Tayler-Spruit instability as well as its occurrence have been questioned by several authors \citep[see][]{maeder2004,denissenkov2007,zahn2007}. Recently, \cite{fuller2019} and \cite{ma2019} re-investigated the Tayler-Spruit mechanism, confuting all previous criticisms to the original work by \cite{spruit2002} and providing an alternative formulation for the angular momentum transport, with efficiency definitively larger than that adopted in the present work.
In this context, we note that in our model including magnetic instabilities the angular velocity profile is flat all along the structure (see lower panel in Fig.~\ref{fig7}), clearly indicating that the adopted formulation for magnetic stresses enforces rigid rotation. As a consequence, the use of more efficient angular momentum transport as proposed by \cite{fuller2019} would not change the conclusion of our work.

As a final consideration, we recall that, when all mass from the sdB-donor is transferred to the accretor, the resulting object is stable, since the angular velocity profile has attained an equilibrium configuration, and it can further evolve only by cooling down.
In Paper I (see Sect. 6 there), we investigated the possibility that the loss of angular momentum by this massive WD in the rotating model could lead to a compressional heating and, eventually, to a new He-burning episode.
For the magnetic model in the present work, we find that the WD, that is the outcome of the CD-30 evolution, cannot emit gravitational waves in the $f$-mode, because the ratio of the rotational energy and the gravitational one of the WD is $\Theta=0.025$.
This is definitively smaller than the critical value for such instabilities to occur.
On the other hand, we find that WD $r$-mode instabilities can arise because their growth timescale is $\tau_r=9\times 10^8$yr, while the viscous timescale $\tau_\nu$ diverges to infinity, as the shear viscosity is negligible along the entire star. Another possibility, as outlined in Paper I, is that a different braking mechanism, currently not identified, could operate on a timescale shorter than $\tau_r$.
As in Paper I, we investigated the latter issue for the magnetic model by computing the further evolution of the massive WD emerging in CD-30 by subtracting at each timestep an amount of angular momentum given by $\delta j=J_{WD}\cdot [1- \exp({-\delta t/\tau})]$, where $\delta t$ is the timestep, $J_{WD}$ the actual value of the total angular momentum of the WD, and $\tau$ varied in the range $1-10^9$ yr.
We found that, independently of $\tau$, He is never ignited.
This conclusion is different from what we obtained for the rotating model presented in Paper I and it is a direct consequence of the very large efficiency of angular momentum redistribution due to the Tayler-Spruit instability. 
In the rotating model, the angular momentum subtraction affects only the most external zones and this determines a local compressional heating which drives an additional He-flash, if the braking timescale is not too long (see the discussion in Paper I, Sect. 6). 
On the contrary, in the magnetic model considered in the present work, the loss of angular momentum implies a global reduction of the rotation velocity along the entire star, which undergoes an homologous compression. This prevents any sizable heating of the He-rich buffer. 

We note that, in our rotating and magnetic models, it is assumed that angular
momentum is deposited onto the WD in such a way that it could attain critical rotation velocity (Eq. (3) in Paper I). The effect of tides that may slow-down rotation awaits exploration.

To summarize, we expect, based on consideration of the effects of rotation and magnetic field-induced instabilities, that neither in CD-30 nor in the similar system \ptf\ a SN~Ia scale explosion due to a double detonation will happen.
We infer that the final configuration for these systems will be a white dwarf with a CO core and a massive He/C/O-envelope accreting matter at a very low rate from the remnant of the sdB component.

\begin{acknowledgements}
	We acknowledge the anonymous referee for carefully reading the manuscript and providing constructive and helpful comments.
	L.P. acknowledges partial financial support from the INAF Minigrant 2023 \textit{Self-consistent Modeling of Interacting Binary Systems} and from the Italian MUR project 2022RJLWHN: \textit{Understanding R-process {\rm \&} Kilonovae Aspects (URKA)}.
	I.D. and E.B. acknowledge partial support from the Spanish project 	PID2021-123110NB-100, financed by MCIN/AEI/10.13039/501100011033/FEDER/UE.
	The work of LRY was carried out under the state assignment to the Institute of Astronomy of RAS.
\end{acknowledgements}

   \bibliographystyle{aa}
   \bibliography{./piersanti}
\begin{appendix}
\onecolumn
\section{Rotating model}\label{a:tables}
\begin{table*}[ht!]
	\caption{Selected properties of the rotating model during the H-accretion phase.}
	\label{test}      
	\centering                   
\begin{tabular}{r c c c c c c c c }
		\hline\hline                 
\# Fl. & $\Delta T$ & $\rho_{ig}$ & $T_H^{max}$ & $L_H^{max}$ & $\Delta M_{acc}$ & $\Delta M_{lost}$ & <\mdote> & $\eta$ \\
	\hline                 
  1    &      2.60  &     4.53    &     1.66    &   388.97    &      0.63        &       1.84        &   2.42   & -1.92  \\
  2    &      2.27  &     4.75    &     1.68    &   353.67    &      1.02        &       2.51        &   4.50   & -1.45  \\
  3    &      1.80  &     5.93    &     1.81    &  2465.40    &      0.68        &       2.82        &   3.79   & -3.13  \\
  4    &      1.21  &     3.75    &     1.60    &   101.35    &      0.37        &       1.63        &   3.07   & -3.35  \\
  5    &      0.77  &     2.00    &     1.33    &    29.82    &      0.21        &       0.61        &   2.68   & -1.97  \\
  6    &      0.64  &     1.79    &     1.31    &    20.89    &      0.16        &       0.56        &   2.48   & -2.58  \\
  7    &      0.67  &     1.75    &     1.30    &    20.48    &      0.16        &       0.52        &   2.32   & -2.34  \\
  8    &      0.63  &     1.69    &     1.29    &    15.81    &      0.14        &       0.50        &   2.23   & -2.56  \\
  9    &      0.65  &     1.71    &     1.30    &    17.42    &      0.14        &       0.50        &   2.16   & -2.57  \\
 10    &      0.67  &     1.74    &     1.31    &    18.23    &      0.14        &       0.50        &   2.12   & -2.53  \\
 11    &      0.67  &     1.77    &     1.31    &    20.11    &      0.14        &       0.51        &   2.11   & -2.55  \\
 12    &      0.67  &     1.79    &     1.32    &    21.12    &      0.14        &       0.51        &   2.11   & -2.61  \\
 13    &      0.65  &     1.82    &     1.33    &    22.85    &      0.15        &       0.53        &   2.23   & -2.61  \\
 14    &      0.63  &     1.85    &     1.34    &    26.16    &      0.16        &       0.53        &   2.48   & -2.39  \\
 15    &      0.60  &     1.89    &     1.35    &    26.92    &      0.16        &       0.53        &   2.71   & -2.24  \\
 16    &      0.57  &     1.93    &     1.36    &    26.80    &      0.17        &       0.55        &   2.96   & -2.30  \\
 17    &      0.53  &     1.98    &     1.37    &    29.28    &      0.18        &       0.56        &   3.37   & -2.18  \\
 18    &      0.47  &     2.01    &     1.38    &    30.27    &      0.19        &       0.57        &   3.94   & -2.10  \\
 19    &      0.42  &     2.08    &     1.40    &    31.89    &      0.20        &       0.59        &   4.71   & -1.98  \\
 20    &      0.38  &     2.14    &     1.41    &    35.24    &      0.21        &       0.61        &   5.69   & -1.85  \\
 21    &      0.33  &     2.22    &     1.43    &    38.04    &      0.23        &       0.63        &   7.05   & -1.70  \\
 22    &      0.29  &     2.35    &     1.46    &    41.66    &      0.26        &       0.66        &   9.06   & -1.48  \\
 23    &      0.23  &     2.49    &     1.49    &    46.71    &      0.30        &       0.68        &  13.04   & -1.25  \\
 24    &      0.20  &     2.68    &     1.53    &    51.56    &      0.37        &       0.31        &  18.74   &  0.15  \\
	\hline                 
\end{tabular}
\tablefoot{For each flash episode we report the duration of the accretion phase ($\Delta T$ in Myr), the ignition density ($\rho_{ig}$ in $10^3$\gcc), the maximum temperature attained in the H-burning shell ($T_H^{max}$ in $10^8$ K) and the maximum energy per unit time delivered via H-burning  ($L_H^{max}$ in $10^6$\lsun) during the flash, the mass accreted between two successive flashes ($\Delta M_{acc}$ in $10^{-4}$\,\msun), the mass ejected during the RLOF episode ($\Delta M_{lost}$ in $10^{-4}$\,\msun), the average accretion rate before the onset of RLOF (in $10^{-11}$\msyrm), and the retention efficiency defined as $\eta=1-\Delta M_{lost}/\Delta M_{acc}$.}
\end{table*}
\twocolumn
\end{appendix}

\end{document}